\documentclass[twocolumn,showpacs,preprintnumbers,amsmath,amssymb]{revtex4}

\usepackage{graphicx}
\usepackage{dcolumn}
\usepackage{bm}

\begin{document}

\title{Highly transparent superconducting -- normal junctions in a homogeneous superconductor induced by local fields of magnetic
domains}

\author{J. Fritzsche}
\author{R. B. G. Kramer}
\author{V. V. Moshchalkov}
\affiliation{INPAC - Institute for Nanoscale Physics and
Chemistry, Katholieke Universiteit Leuven, Celestijnenlaan 200 D,
B-3001 Leuven, Belgium}

\date{\today}

\begin{abstract}
Using the highly inhomogeneous fields of a magnetic substrate,
tunable junctions between superconducting and normal state regions
were created inside a thin film superconductor. The investigation
of these junctions, created \emph{in the same material}, gave
evidence for the occurrence of Andreev reflection, indicating the
high transparency of interfaces between superconducting and normal
state regions. For the realization of this study, a ferromagnet
with magnetic stripe domains was used as a substrate, on top of
which a superconducting transport bridge was prepared
perpendicular to the underlying domains. The particular choice of
materials allowed to restrict the nucleation of superconductivity
to regions above either reverse-domains or domain walls. Moreover,
due to the specific design of the sample, transport currents in
the superconductor passed through a sequence of normal and
superconducting regions.
\end{abstract}

\pacs{74.45.+c, 75.60.Ch, 74.78.Db}

\maketitle

\section{\label{sec:Introduction}Introduction}

There is a wealth of physical phenomena inherent to junctions
between superconducting (S) and normal (N) matter, such as the
Josephson effect~\cite{Josephson:1974}, quasi-particle
tunneling~\cite{Meservey:1970} and Andreev reflection
(AR)~\cite{Andreev:1964}. Traditionally, SN junctions are either
based on composites of superconducting and normal-conducting
materials, or on constrictions/thickness modulations of
superconducting films. A common feature of these junctions is
their fixed character, since they are static constructions that
can not be modified any more once they are fabricated. However,
recent works have shown that the highly inhomogeneous fields of
ferromagnetic domains can be used to locally suppress
superconductivity in thin S-films, resulting in the states of
domain-wall superconductivity~\cite{Yang:2004} (DWS) or
reverse-domain superconductivity~\cite{Fritzsche:2006} (RDS). In
these two states, the superconductor can be seen as a network of
SN junctions, which can exhibit the same flexibility as the
underlying magnetic domains. In this article, we demonstrate that
tunable SN junctions can be created in a controlled way by using
the highly inhomogeneous fields of ferromagnetic domains. In
particular, we show that in such junctions, the interfaces between
the superconducting and the normal parts are highly transparent
for incident electrons, which is a consequence of creating
superconducting and normal state regions inside the \emph{same
material}.

In order to experimentally investigate SN interfaces that are
induced by stray magnetic fields, a specially designed
superconductor/ferromagnet (S/F)
hybrid~\cite{Buzdin:2005,Lyuksyutov:2005,Aladyshkin:2009} system
was needed, exhibiting the following two qualities: (i) The
opportunity to specifically realize superconductivity either above
the magnetic domain walls (DWS) or above the reverse domains (RDS)
of the substrate. (ii) Transport currents had to cross effectively
the interfaces between the induced superconducting and
normal-state regions. The preparation of such system is
challenging as several strict requirements need to be fulfilled.
First of all, formation of magnetic stripe domains in the template
is desirable~\cite{Belkin:2008}. Alignment of a transport bridge
perpendicular to such domains guarantees a bias current to cross
them successively. Second, the magnetic domain pattern of the
template must not change significantly when subjected to external
fields, required for setting up the different states (e.g. RDS) in
the S-layer. Finally, the out-of plane component of the stray
field above magnetic domains has to reach the upper critical field
of the superconductor \cite{Sonin:1988}. Thereby, realization of
DWS is possible down to temperatures well below $T_\mathrm{c}$.
\begin{figure}[b]
\includegraphics[width=86mm]{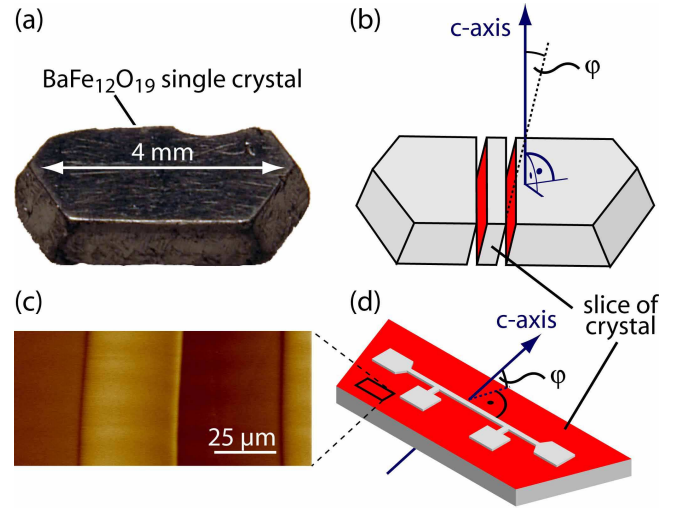}
\caption{\label{fig: sample preparation} (Color online) (a)
Photograph of the as-grown barium hexaferrite single crystal. (b)
Sketch of the crystal after slicing under a cut angle $\varphi$
with respect to its c-axis. (c) Magnetic force microscopical image
of the magnetic domains at the cut surface of a slice of the
BaFe$_{12}$O$_{19}$ single crystal. (d) Sketch of a transport
bridge processed perpendicular to the c-axis on the cut surface of
a slice of the crystal.}
\end{figure}

\section{\label{sec:Experimental}Experimental}

Accounting for the above requirements, S/F hybrid systems were
prepared by slicing a single crystal of barium hexaferrite
(BaFe$_{12}$O$_{19}$) under a small tilt to its c-axis
(Figure~\ref{fig: sample preparation}(b)), and then processing
superconducting aluminium bridges of 50~nm thickness on the cut
surfaces perpendicular to the c-axis (Figure~\ref{fig: sample
preparation}(d)). The superconducting and ferromagnetic components
were electrically isolated by 5~nm SiO$_2$ in order to prevent any
proximity effect. The ferromagnetic crystal (Figure~\ref{fig:
sample preparation}(a)) was grown from a sodium carbonate flux,
following a recipe after \cite{Gambino:1961}. When cut along the
proper crystallographic axis, single crystals of
BaFe$_{12}$O$_{19}$ exhibit a one-dimensional stripe-type domain
structure (Figure~\ref{fig: sample preparation}(c)) with dominant
in-plane magnetization and relatively small out-of-plane component
$M_\mathrm{z}$ \cite{Hubert:403}.

\begin{figure}[b]
\includegraphics[width=67mm]{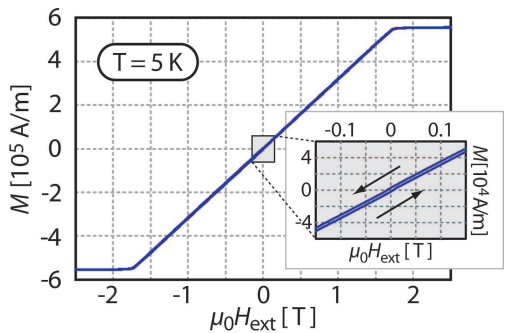}
\caption{\label{fig: M(H)} (Color online) The magnetization $M$ of
a slice of the BaFe$_{12}$O$_{19}$ single crystal at 5~K as a
function of the external magnetic field $\vec{H}_\mathrm{ext}$
normal to the cut surface. The insert shows a magnification of the
curve for $|H_\mathrm{ext}|\leq$ 150~mT.}
\end{figure}

To demonstrate that these magnetic domains do not change
significantly in perpendicular external magnetic fields
$|H_\mathrm{ext}|\leq120$ mT, the magnetization $M$ of one slice
of the single crystal was measured with a vibrating sample
magnetrometer as a function of $H_\mathrm{ext}$ (see
Figure~\ref{fig: M(H)}). Apparently, the magnetization of the
ferromagnet depends almost linearly on the perpendicular applied
magnetic field and saturates at $H_\mathrm{ext}\simeq$ 1.7~T. From
the slope $dM/dH\sim3.2\cdot10^5$ Am$^{-1}$T$^{-1}$, one can
indeed expect only minor changes of the domain structure for
$|H_\mathrm{ext}|\leq120$ mT, since the corresponding variation of
the magnetic moment is less than 7$\%$ of the saturated
magnetization ($5.5\cdot10^5$ Am$^{-1}$T$^{-1}$).
Furthermore, the influence of the external magnetic field on the
size and position of the magnetic domains was studied at
low-temperatures (77~K) with a scanning Hall-Probe
microscope~\cite{Bending:1999}. As it is shown in Figure~\ref{fig:
domain width}, the width $w$ of the parallel domains increases
linearly for $H_\mathrm{ext}\leq150$~mT with a rate of
approximately 43~nm/mT. In accordance with the very small
coercivity of these ferromagnets (see Figure~\ref{fig: M(H)}), the
observed domain walls returned to their initial positions within
the experimental resolution of 1~$\mu$m, each time
$H_\mathrm{ext}$ was reduced to zero.
\begin{figure}[t]
\includegraphics[width=72.5mm]{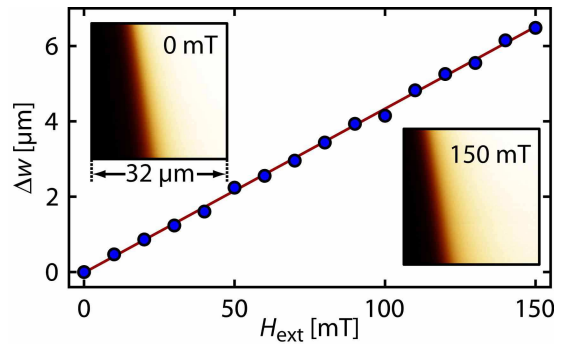}
\caption{\label{fig: domain width} (Color online) The increase of
the width $w$ of a parallel domain (the bright areas in the
inserts) in the ferromagnetic substrate as a function of
$H_\mathrm{ext}$. The shown data is the average of the results
obtained for an increasing and a decreasing external magnetic
field. The inserts show two examples of the position of the same
domain-wall at 0~mT and 150~mT, respectively.}
\end{figure}

\begin{figure}[t]
\includegraphics[width=86mm]{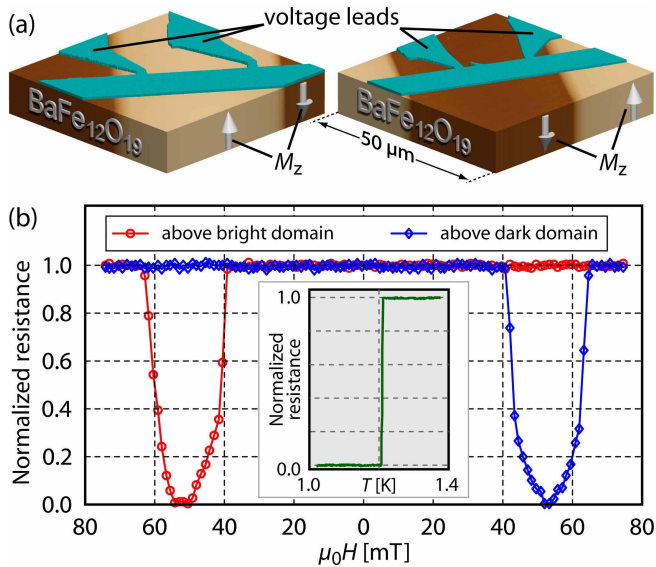}
\caption{\label{fig: RDS} (Color online) (a) AFM-images of two
transport bridges with their underlying magnetic domains
(MFM-images). The MFM-images are vertically extended to illustrate
the domains. Arrows indicate the z-component $M_\text{z}$ of the
magnetization of the template. (b) The normalized resistance of
the two bridges, measured at 340~mK with a bias current of
10~$\mu$A as a function of $H_\mathrm{ext}$. The insert shows the
superconducting transition of a reference Al-film on a
Si-substrate as a function of temperature (critical current
density $j_\mathrm{c}\sim1.2\cdot10^8$ A/m$^2$ at 340~mK).}
\end{figure}

In order to show that the prepared S/F hybrid systems are suitable
for the study of stray field induced SN interfaces,
Figure~\ref{fig: RDS}(b) shows the normalized resistance
$R^\ast:=R/R_\mathrm{N}$ ($R_\mathrm{N}$ being the normal state
resistance) of two transport bridges as a function of
$H_\mathrm{ext}$. Both curves were measured at 340~mK, i.e. well
below the critical temperature $T_\mathrm{c}\simeq1.2$ K of the
used aluminum (see the insert of Figure~\ref{fig: RDS}). From the
corresponding atomic and magnetic force microscopical images (AFM
and MFM, respectively) of Figure~\ref{fig: RDS}(a), it can be seen
that the measured parts of the bridges lay entirely above magnetic
domains of opposite magnetization. The difference in the MFM
signal above the two kinds of domains indicates a non-zero out-of
plane component $B_\mathrm{stray}^\mathrm{z}$ of the stray
magnetic field. Note that $B_\mathrm{stray}^\mathrm{z}\neq 0$
above the wide domains results directly from a finite cutting
angle $\varphi$ to the c-axis of the crystal. Therefore, by
choosing $\varphi$, the strength of $B_\mathrm{stray}^\mathrm{z}$
can be adapted to match the critical fields of the superconductor.
For the present case, it was found that aluminium as a
superconductor and $\varphi=10^\circ$ are a good match. For the
case of the bridge above the bright domain (left panel of
Figure~\ref{fig: RDS}(a)), $R^\ast$ drops to zero around -53~mT
(see the red curve with circles). In a symmetric manner, $R^\ast$
of the bridge above the dark domain shows a similar behavior
around +53~mT (see the blue curve with diamonds). These
observations prove the possibility to realize the state of RDS by
applying compensation fields of $H_\mathrm{ext}=\pm53$~mT to the
designed Al/BaFe$_{12}$O$_{19}$ hybrids. Moreover, it becomes
clear that in the state of RDS at 340~mK, the superconducting
order parameter is completely suppressed above the corresponding
parallel domains (i.e. above magnetic domains with magnetization
in the same direction as $\vec{H}_\mathrm{ext}$).

Next, a long transport bridge of $35\times 250$ $\mu m^2$ was
investigated, which -- due to its relatively large size -- had to
cross several magnetic domain walls of the substrate. Inspection
of the sample with a magnetic force microscope revealed indeed
nine domain walls underneath the bridge (Figure~\ref{fig:
fig_1}(a)). Figure~\ref{fig: fig_2} displays the normalized
dc-resistance of the bridge well below $T_\mathrm{c}$, measured as
a function of bias current $I$ and $H_\mathrm{ext}$. As can be
expected from the above presented results, two pronounced minima
in resistivity are seen around $\pm53$~mT, indicating that stray
fields above magnetic domains are compensated by $H_\mathrm{ext}$.
Application of these compensation fields thus induces the RDS
state in the S/F hybrid system. Furthermore, a second key feature
can be seen in Figure~\ref{fig: fig_2}. While beyond the
compensation fields the resistance quickly rises towards its value
in the normal state, parts of the bridge remain superconducting
when subjected to external fields lower than the compensation
fields. Particularly, in the case of zero applied field, when
superconductivity is likewise suppressed above domains of opposite
magnetization, the reduced resistance is a clear fingerprint of
DWS. Moreover, the insert in Figure~\ref{fig: fig_2} shows the
transitions of the bridge from the normal state to the states of
DWS and RDS (at 0~mT and 53~mT, respectively) as a function of
temperature. The significant difference $\Delta T_\mathrm{c}$ of
the onsets of the transitions reflects the confinement of the
superconducting order parameter above wide magnetic domains (RDS)
and narrow domain-walls (DWS) \cite{Aladyshkin:2006}.
\begin{figure}[b]
\includegraphics[width=86mm]{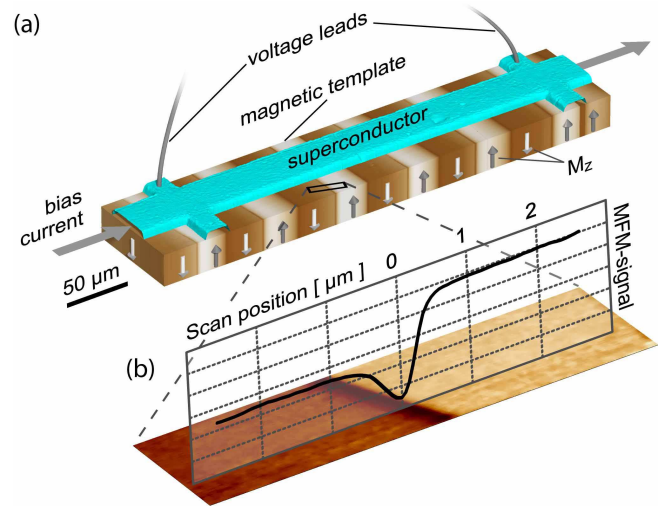}
\caption{\label{fig: fig_1} (Color online) (a), The transport
bridge (optical image shown in three dimensions) with its
underlying magnetic domains (MFM image). The MFM image is
vertically extended to illustrate the domains. Arrows indicate the
z-component $M_\text{z}$ of the magnetization of the template.
(b), A detailed MFM-image of a typical domain wall in the
substrate.}
\end{figure}
\begin{figure}[t]
\includegraphics[width=86mm]{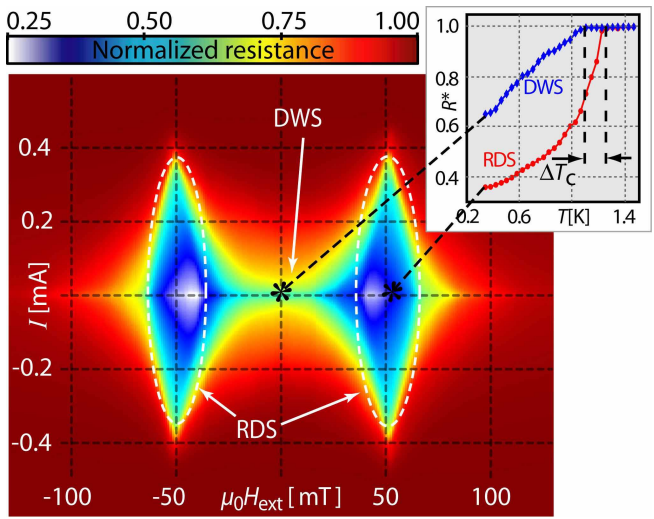}
\caption{\label{fig: fig_2} (color online) The normalized
resistance of the transport bridge of Figure~\ref{fig: fig_1}(a)
at 340 mK, as a function of bias current $I$ and external field
$H_\text{ext}$ ($R_\text{N}=3.62~\Omega$). Regions where RDS and
DWS occur are indicated. The insert shows the resistive
transitions ($I$=10 $\mu$A) from the normal state to the states of
DWS (blue curve with diamonds) and RDS (red curve with circles).
The points corresponding to $T$=340~mK are indicated in the main
panel.}
\end{figure}

Above, the occurrence of the minima in Figure~\ref{fig: fig_2}(a)
has been discussed, along with the reduction of resistance at zero
applied field. However, surprising are the values of the
resistance reached at these points: As can be seen from
Figure~\ref{fig: fig_1}(a), approximately half of the area of the
bridge is covered by each kind of domains. Nucleation of
superconductivity above one type of domains should thus cause the
bridge to loose roughly half of its resistance in the normal
state. By contrast, for compensation fields of both polarities,
only half of the expected resistance is seen.

A similar observation can be made at $H_\text{ext}=0$ when
superconductivity survives above domain walls only. In that case,
the drop in resistance is, a priori, expected to be equal to the
ratio between the width of magnetic domains and domain walls. But
from detailed MFM studies it becomes clear that all changes in
stray fields are confined to approximately 1 $\mu m$ around domain
walls (Figure~\ref{fig: fig_1}(b)), whereas the domains are
typically 25 $\mu m$ wide (Figure~\ref{fig: fig_1}(a)). Therefore,
in absence of external fields, the observed reduction of
resistance by $\sim 45$\% is surprising.

In order to investigate these remarkable features in more detail,
the differential resistance $dV/dI$ of the transport bridge was
measured as a function of bias current and temperature in both
states, RDS and DWS. Simultaneously, the voltage drop $V$ over the
bridge was also detected. Measurements were carried out via
standard lock-in techniques at a frequency of 33 Hz and an
ac-modulation current of 2 $\mu$A. The normalized differential
conductance $G_\text{N}=\frac{dI}{dV}R_\text{N}$ is shown in
Figure~\ref{fig: fig_3} as a function of voltage for
$H_\text{ext}=\pm53$ mT (RDS). In that diagram, results are shown
twice for clarity: the left 2D-panel displays a few conductance
curves that are vertically shifted, whereas all obtained curves
are given in a 3D-representation at the right. Here, at lowest
temperatures, $G_\text{N}$ is sharply peaked at zero voltage,
declining symmetrically to its minima at $\pm V_1$ before
recovering to its normal value at higher voltages. Together with
some smaller local minima, these features gradually collapse with
increasing temperature.

\section{\label{sec:Discussion}Discussion}

In order to interpret the conductance spectra of Figure~\ref{fig:
fig_3}, several aspects must be taken into account:

(i) The whole transport bridge is in the normal state for
$|V|>$2~mV even at the lowest temperature (340~mK). The reason for
this is that the critical current density $j_c$ is exceeded due to
the low resistance of the bridge. Therefore, in a certain
low-voltage region where $j<j_c$, a higher value for $G_\text{N}$
is expected, since the parts of the bridge above reverse-domains
(RD) are superconducting.

(ii) The normalized differential conductance reaches 2.8 at 340~mK
and zero voltage. Assuming that this increase of $G_\mathrm{N}$
was solely caused by the N~$\rightarrow$~S transition mentioned
under point (i), approximately 64\% of the transport bridge had to
become superconducting. However, the hybrid system behaves similar
for both polarities of $H_{\mathrm{ext}}$ (see Figure~\ref{fig:
fig_2}), meaning that an unequal distribution of parallel and
reverse domains can not be the reason for the high conductances
observed at positive \emph{and} negative compensation fields.
Moreover, as discussed above, the external field increases the
width of the parallel domains by 43~nm/mT (see Figure~\ref{fig:
domain width}). Accordingly, at 53~mT, a normalized conductance of
only 1.8 instead of 2 should be expected, provided that parallel
and reverse domains are equally distributed at $H_\mathrm{ext}=0$.
Finally, the characteristic length $\xi_\text{N}=\sqrt{\hbar
D/k_\mathrm{B}T}$, over which the Cooper pair amplitude decays
exponentially with the distance from an SN interface, is in the
present case of the order of 400~nm \footnote{An estimation of the
diffusion coefficient is
$D=v_\mathrm{F}l_\mathrm{p}/3\approx6.6\cdot10^{-3}$ m$^2$/s, with
the Fermi velocity $v_\mathrm{F}=2.03\cdot 10^6$ m/s
\cite{Ashcroft} and the electron mean free path
$l_\mathrm{p}=mv_\mathrm{F}/\varrho_\mathrm{i}ne^2\approx9.8$ nm
($m$ and $e$ are electronic mass and charge, $n=18.1\cdot 10^{28}$
m$^{-3}$ is the density of conduction electrons \cite{Ashcroft}).
The residual resistance $\varrho_i$ was estimated according to
$\varrho_\mathrm{p}/\varrho_\mathrm{i}+1\approx R(\text{295
K})/R(\text{1.5 K})= 1.68$, with the resistivity
$\varrho_\mathrm{p}=2.74\cdot10^{-8}$ $\Omega$m of Al at 295 K
\cite{Kittel}.}. Due to this proximity effect, the superconducting
state extends into the normal regions and vice versa, but the
corresponding increase of $G_\text{N}$ at $V=0$ is only minor.

\begin{figure}[b]
\includegraphics[width=86mm]{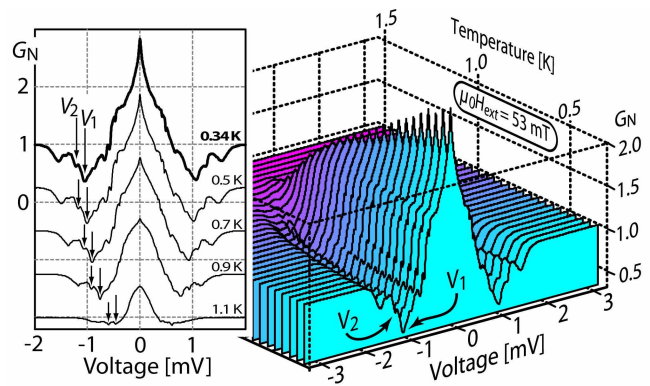}
\caption{\label{fig: fig_3} (color online) Differential
conductance spectra of the transport bridge of Figure~\ref{fig:
fig_1}(a) at 53 mT in the state of RDS. Arrows mark two minima
whose positions can be traced. In the left panel, curves are
shifted for clarity and the vertical scale corresponds to the top
curve.}
\end{figure}

Taking account of the above considerations, the observed
conductance of 2.8 can not be explained by a corresponding
expansion of the superconducting state along the transport bridge.
However, below the superconducting gap $\Delta$ (i.e. for
$V<\Delta/e$), an excess of the conductance can generally result
from Andreev reflection processes at SN-interfaces, if the latter
are highly transparent for incident electrons. In the present
case, normal and superconducting states are created \emph{inside
the same material} and, therefore, the presence of higly
transparent SN-interfaces is reasonable. Accordingly, the observed
excess of conductance suggests that the mechanism of charge
transfer across the SN-interfaces is affected by AR.

The theory of Blonder, Tinkham and Klapwijk (BTK)
\cite{Blonder:1982} describes the effects of AR on the conductance
of a single SN junction for the particular case of ballistic
transport in the normal-state region. From that theory it follows
that inside the gap, the conductance can be enhanced up to twice
its above-gap value. In the present case, as discussed above, the
above-gap conductance in the RDS-state at 53~mT can be estimated
to be 1.8. Therefore, the observed zero-voltage conductance of 2.8
is smaller than twice the above-gap conductance ($2\times1.8$),
meaning that these findings are not in contradiction with the
BTK-theory.

Moreover, the BTK-theory predicts for highly transparent
interfaces a flat conductance below the gap, which has been
verified experimentally with superconducting point contacts (see
for example \cite{Soulen:1998}). By contrast, the
$G_\mathrm{N}$-curves of Figure~\ref{fig: fig_3} are sharply
peaked at zero voltage. Such anomalies in the conductance spectra
in the form of zero-bias peaks have been reported before in
systems that deviate from the model of BTK, such as for example
planar Nb/Au contacts~\cite{Xiong:1993}, junctions between
superconductors and
semiconductors~\cite{Kleinsasser:1990,Nguyen:1992,Kastalsky:1991}
and series of SNS-junctions~\cite{Kvon:2000}. In the present case,
the used sample differs also significantly from the model system
of BTK, since the bridge crosses nine domain walls (see
Figure~\ref{fig: fig_1}(a)), each of them inducing one SN
interface. Moreover, due to the large size of the domains, the
electric transport in the normal-state regions is not ballistic. A
theoretical description of such series of diffusive SNS junctions
will go beyond the ballistic
theories~\cite{Blonder:1982,Octavio:1983}, and will have to
include nonlocal coherent effects in the normal-state regions
\cite{Wees:1992,Nazarov:1996}.

(iii)
\begin{figure}[b]
\includegraphics[width=86mm]{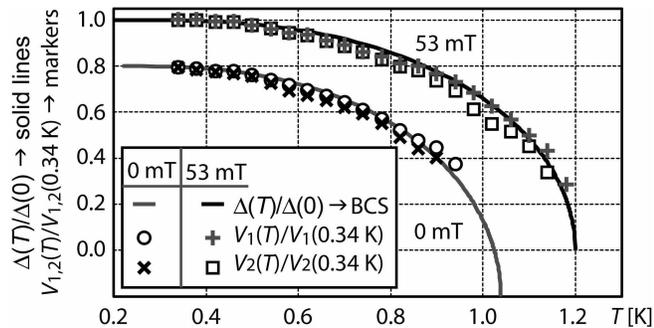}
\caption{\label{fig: fig_4} The positions of characteristic minima
in the conductance spectra of Figure~\ref{fig: fig_3} and
\ref{fig: fig_5} (markers) are compared with $\Delta(T)$ (solid
lines). For clarity, the results obtained at 0 mT are shifted by
-0.2\ .}
\end{figure}
Two of the local minima of the conductance spectra of
Figure~\ref{fig: fig_3}, marked as $V_1$ and $V_2$, can be traced
from 340~mK to nearly $T_\mathrm{c}$. Their relative position on
the $V$-axis was compared to the superconducting gap function
\begin{equation*}
    \ln\left(\cfrac{1.13E_\mathrm{c}}{k_\mathrm{B}T_\mathrm{c}}\right)=\int_0^{E_\mathrm{c}}\cfrac{\tanh\left(0.5k_\mathrm{B}^{-1}T^{-1}\sqrt{\xi^2+\Delta^2}\right)}{\sqrt{\xi^2+\Delta^2}}\ d\xi,
\end{equation*}
of the BCS-theory~\cite{Tinkham:63}, using a value of 423 K for
the ratio between cut-off frequency $E_\mathrm{c}$ and Boltzmann
constant $k_\mathrm{B}$~\cite{Gschneidner:1964}. A solution
$\Delta(T)$ of the above equation can be found by iteration,
integrating numerically over energies $\xi$ while treating
$T_\mathrm{c}$ as a fitting parameter. As illustrated in
Figure~\ref{fig: fig_4} (upper curve), $V_{1,2}$ follow strictly
the superconducting gap $\Delta$ in temperature.

For the ideal case of a single ballistic SNS junction, it is known
that multiple Andreev reflection (MAR) leads to minima in the
conductance curves at voltages smaller than the gap
($V<\Delta/e$). Their positions follow $\Delta(T)$ in the same way
as $V_{1,2}$. However, the present case is quite different in that
the measured $G(V)$-curves belong to a series of diffusive SNS
junctions. When dividing $V$ by the number of SN
interfaces~\cite{Baturina:2002} and considering that $V$ dropped
mainly over the normal-state regions of the bridge, it could be
concluded that $V_{1,2}$ lay inside the gap and originate from MAR
(typical values for $\Delta/e$ are 200 $\mu$V for
Al~\cite{Giaever:1961}). However, it is also possible that series
of AR processes lead to multiplication effects and to different
effective voltages across subsequent SN interfaces. Therefore,
even if caused by the same process, features in $G(V)$ could
repeatedly appear at different voltages, and result in the
observed set of local minima. Moreover, a multiplication effect in
series of junctions might also lead to an increase of the
conduction by factors higher than two~\cite{Shan:2003}.
\\\\
\begin{figure}
\includegraphics[width=86mm]{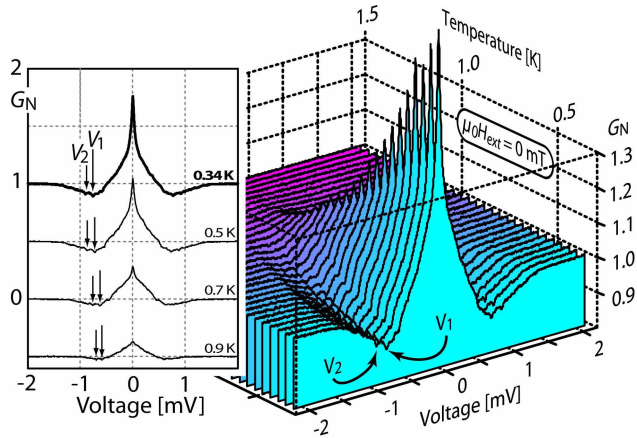}
\caption{\label{fig: fig_5} (color online) Differential
conductance spectra of the transport bridge of Figure~\ref{fig:
fig_1}(a) at 0 mT (DWS). Arrows mark two minima whose positions
can be traced. In the left panel, curves are shifted for clarity
and the vertical scale corresponds to the top curve.}
\end{figure}
Intriguingly, all observations described above can be made not
only in the case of RDS but also in absence of external fields
when DWS is realized (Figure~\ref{fig: fig_5}). In that case,
local minima are less pronounced, but nonetheless, two of them can
be traced up to higher temperatures (lower curve in
Figure~\ref{fig: fig_4}). As before, their positions in the
conductance spectra follow the collapse of $\Delta$. It is
remarkable that values of $T_\mathrm{c}$ obtained by fitting are
significantly different in the cases of RDS ($1.20\pm0.02$~K) and
DWS ($1.05\pm0.03$~K). These findings reflect directly that due to
quantum size effects, $T_\mathrm{c}$ values of superconducting
micro-structures differ significantly from those of bulk
superconductors~\cite{Moshchalkov:1995} -- an effect that leads to
the reduction of $T_\mathrm{c}$ when superconductivity is confined
above the domain walls of a underlying
ferromagnet~\cite{Gillijns:2007PRB76}.

\section{\label{sec:Conclusions}Conclusions}
In conclusion, there are two major findings of this work: On one
hand, it has been demonstrated that tunable SN junctions inside
superconducting thin films can be created \emph{in a controlled
manner} by using magnetic templates. This first conclusion is a
direct result from the successful fabrication of a S/F hybrid
system, that allows for setting up DWS and RDS in the S-layer,
without changing the actual configuration of magnetic domains. On
the other hand, the occurrence of Andreev reflection, observed in
the conductance of the S-layer of the hybrid system, proves the
high transparency of SN interfaces induced by magnetic stray
fields. This result is based on the innovative approach to create
SN junctions \emph{in the same material} via local suppression of
superconductivity.

From a technological point of view, generation of SN junctions via
ferromagnets is attractive due to both, the natural tunability of
magnetic domain structures and the here demonstrated high quality
of SN interfaces. Potentially, inclusion of magnetic templates
with pure in-plane magnetization will make it possible to invert
the scheme of DWS and to \emph{suppress} superconductivity in a
very narrow region above domain walls, realizing the domain-wall
normal state (DWN). Such configuration may lead to controllable
phase coupling effects between two superconducting reservoirs,
separated by a thin DWN region, and pave the way for the
development of new types of tunable quantum interference devices.

\begin{acknowledgments}
This work is supported by the FWO, GOA and IAP projects and the
ESF-NES Research Networking Programme.
\end{acknowledgments}


\end{document}